\documentclass[twocolumn,showpacs,aps,prl,superscriptaddress,preprintnumbers,amsmath,amssymb,floatfix]{revtex4}

\usepackage{graphicx}
\usepackage{dcolumn}
\usepackage{bm}

\usepackage{relsize}
\usepackage{xspace}
\def\babar{\mbox{\slshape B\kern-0.1em{\smaller A}\kern-0.1em
    B\kern-0.1em{\smaller A\kern-0.2em R}}}
\def\pep2{PEP-II}

\def\Km{K^-}
\def\Kp{K^+}
\def\piz{\pi^0}
\def\pip{\pi^+}

\def\Kbar{\kern 0.2em\overline{\kern -0.2em K}{}\xspace}

\newcommand{\gevc}{\ensuremath{{\mathrm{\,Ge\kern -0.1em V\!/}c}}\xspace}
\newcommand{\mevc}{\ensuremath{{\mathrm{\,Me\kern -0.1em V\!/}c}}\xspace}
\newcommand{\gevcc}{\ensuremath{{\mathrm{\,Ge\kern -0.1em V\!/}c^2}}\xspace}
\newcommand{\mevcc}{\ensuremath{{\mathrm{\,Me\kern -0.1em V\!/}c^2}}\xspace}

\newcommand{\BABARPubYear}    {06}
\newcommand{\BABARPubNumber}  {044}
\newcommand{\SLACPubNumber} {11993}

\begin{document}
\vspace{1cm}
\begin{flushleft}
\preprint{BABAR-PUB-\BABARPubYear/\BABARPubNumber} 
\preprint{SLAC-PUB-\SLACPubNumber} 
\end{flushleft}
\title{\boldmath Observation of a New $D_s$ Meson Decaying to
$D K$ at a Mass of 2.86~\gevcc}

%
\author{B.~Aubert}
\author{R.~Barate}
\author{M.~Bona}
\author{D.~Boutigny}
\author{F.~Couderc}
\author{Y.~Karyotakis}
\author{J.~P.~Lees}
\author{V.~Poireau}
\author{V.~Tisserand}
\author{A.~Zghiche}
\affiliation{Laboratoire de Physique des Particules, F-74941 Annecy-le-Vieux, France }
% ...
\collaboration{The \babar\ Collaboration}
\noaffiliation

\date{\today}

\begin{abstract}
We observe a new $D_s$ meson with mass
$(2856.6\pm 1.5_{stat.} \pm 5.0_{syst.})$ \mevcc and width $(48 \pm 7_{stat.} \pm 10_{syst.}) \ \mevcc$ decaying into $D^0\Kp$ and $D^+K^0_S$. In the same mass distributions
we also observe a broad structure
with mass 
$(2688 \pm 4_{stat.}  \pm 3_{syst.})$ \  \mevcc and width
$(112 \pm 7_{stat.} \pm 36_{syst.})$ \mevcc.
To obtain this result we use 240~${\rm fb}^{-1}$ of data
recorded by the \babar\  detector at the \pep2 asymmetric-energy $e^+e^-$
storage rings at the Stanford Linear Accelerator Center
 running at center-of-mass energies near 10.6 GeV.
\end{abstract}


\maketitle
The spectrum of known $c\overline{s}$ states can be described economically as 
two s-wave states ($D_s^+$, $D_s^{*+}$) with $J^{P}=0^-,1^-$, and four p-wave 
states ($D^*_{s0}(2317)^+$, $D_{s1}(2460)^+$, $D_{s1}(2536)^+$, $D_{s2}(2573)^+$) with 
$J^P=0^+,1^+,1^+,2^+$, though the last two spin-parity assignments are not 
firmly established.  Whether this picture is correct remains controversial 
because the states at 2317 \mevcc  and 2460 \mevcc~\cite{dsnew} had been expected to lie at much 
higher masses~\cite{th}.

We report here on a new $c\overline{s}$ state and a broad structure 
observed in the decay channels $D^0\Kp$ 
and $D^+K^0_S$.
This analysis is based on
 a 240~${\rm fb}^{-1}$ data sample recorded near the
$\Upsilon(4S)$ resonance by the \babar\ detector at the \pep2
asymmetric-energy $e^+e^-$ storage rings.  

The \babar\ detector is
described in detail elsewhere~\cite{babar}.
Charged particles are detected
and their momenta measured by a combination of a cylindrical drift chamber (DCH)
and a silicon vertex tracker (SVT), both operating within a
1.5 T solenoidal magnetic field. 
A ring-imaging Cherenkov detector (DIRC) is used for
charged-particle identification. Photon energies are measured with a 
CsI electromagnetic calorimeter.
We use information from the DIRC and energy-loss measurements in the 
SVT and DCH to identify charged kaon and pion candidates.

We observe three inclusive processes~\cite{footnote}:
\begin{eqnarray}
e^+ e^- &\to& D^0 K^+ X, D^0 \to K^- \pi^+\\
e^+ e^- &\to& D^0 K^+ X, D^0 \to K^- \pi^+ \pi^0\\
e^+ e^- &\to& D^+ K^0_S X, D^+ \to K^- \pi^+ \pi^+, K^0_S \to \pi^+ \pi^-
\end{eqnarray}

For channels (1) and (2) we perform a vertex fit for the $K^-\pi^+$
and require a $\chi^2$ probability greater than 0.1\%. For the $\pi^0$ in channel (2), 
we consider the photons
that emanate from the $K^-\pi^+$ vertex, perform a fit with 
the $\pi^0$ mass constraint, and require a $\chi^2$ probability greater than 1\%. The combinatorial background is reduced by requiring the $\pi^0$ laboratory momentum
to be greater than 350 MeV/$c$. 

To purify the $D^0$ sample in channel (2), its quasi-two body decays~\cite{cleo}
$K^*\pi$ and $K\rho$ are used, allowing ranges of $\pm 50$
\mevcc around the $K^*$ mass for $K \pi$ and
$\pm 100$ \mevcc around the $\rho$ mass for $\pi \pi$.

\begin{figure}
\vskip -0.15in
\includegraphics[width=0.8\linewidth]{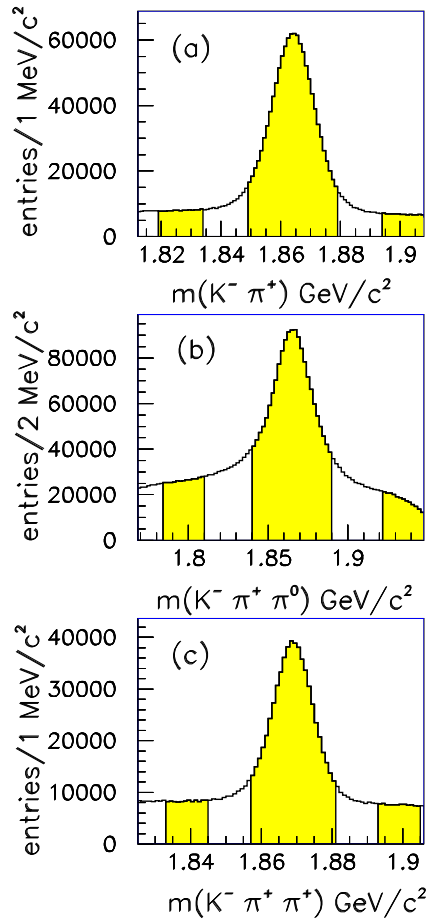}
\caption{\label{fig:fig1}
(a) $\Km\pip$, (b) $\Km\pip\piz$ and (c) $\Km\pip\pip$ 
mass distributions for all candidate events to channels (1), (2) and (3)
respectively. The shaded regions indicate the definition of signal and sidebands 
regions.
}
\end{figure}

\begin{figure*}
\vskip -0.15in
\includegraphics[width=1.0\linewidth]{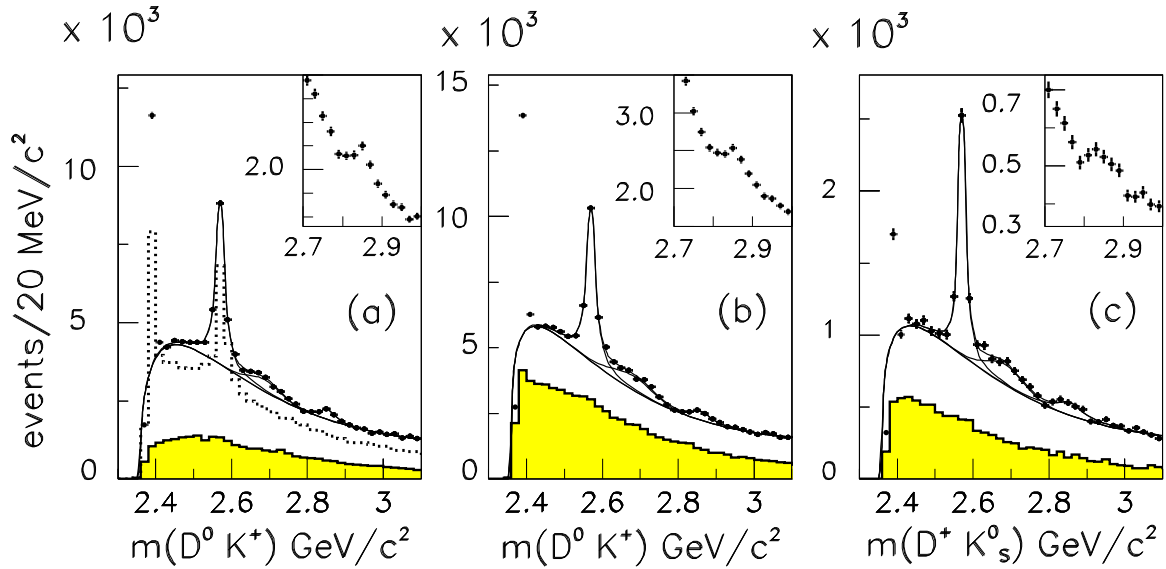}
\vskip -0.15in
\caption{\label{fig:fig2} 
The $D K$ invariant mass distributions for (a) $D^0_{K^-\pi^+} K^+$,
(b) $D^0_{K^-\pi^+\pi^0} K^+$ and (c) $D^+_{K^-\pi^+\pi^+} K^0_s$.
The shaded histograms are for the $D$-mass sideband regions.
The dotted histogram in (a) is from $e^+ e^- \to c \bar c$ Monte Carlo 
simulations
incorporating previously known $D_s$ states with an arbitrary normalization. 
The insets show an expanded view 
of the 2.86 \gevcc region. The solid curves are the fitted background threshold
functions from the three separate fits
described in the text.
}
\end{figure*}

For channel (3), we fit two pions with the same charge and a kaon of the
opposite charge to a common vertex to form the $D^+$ candidate, and require a 
$\chi^2$ probability greater than 0.1\%.  We
obtain the $K^0_S$ sample with a fit that constrains the mass and
require a $\chi^2$ probability greater than 2\%.  $K_S^0$ candidates are retained only if
their
decay lengths are greater than 0.5 cm. 

For all three channels, the $D$ candidate is combined 
with an identified $K$, requiring a vertex fit $\chi^2$ probability
greater than 0.1\%, and constraining the vertex to be in the $e^+e^-$ 
luminous region. 
To reduce combinatorial background from
the continuum ($e^+ e^- \to q \bar q, q=u,d,s,c$) and $B$-meson decays, each
$D K$ candidate must have a momentum $p^*$ in 
the $e^+e^-$ center-of-mass frame greater than 3.5~\gevc.

Figs.~\ref{fig:fig1}(a), (b) and (c) show the $\Km\pip$, $K^-\pi^+\pi^0$, and 
$K^-\pi^+\pi^+$ invariant mass distributions,
respectively.
All distributions show pronounced peaks 
at the $D$ mass, with signal yields of about 950,000, 790,000, and 430,000
events.  Fits using a polynomial and a single Gaussian give $\sigma = 7.6, 12.6, 6.0$ \mevcc 
for the three widths.  
We define the signal region by $\pm 2\sigma$ from the fitted $D$ mass and establish 
sidebands at $(-6\sigma, -4\sigma)$ and $(4\sigma, 6\sigma)$. In the
signal regions, the signal-to-background ratios are 4.1, 1.2, and 2.2 respectively.

Selecting events in the $D$ signal regions, Fig.~\ref{fig:fig2} shows the 
$D^0 K^+$ invariant mass
distributions for channels (1) and (2), and the $D^+ K^0_S$ invariant
mass distribution for channel (3).  To improve mass
resolution, the nominal $D$ mass and the reconstructed 3-momentum
are used to calculate the $D$ energy for channels (1) and (3).
Since channel (2) has a poorer $D^0$ resolution, each $\Km \pi^+ \pi^0$
candidate is kinematically fit with a $D^0$ mass constraint and
we require a $\chi^2$ probability greater than 0.1\%.

We find that the fraction of events
having more than one $DK$ combination per event is 0.9\% for 
channels (1) and (3) and 3.4\% for channel (2).
In the rest of the paper, we use the term reflection to describe enhancements produced 
by two or three body 
decays of 
narrow resonances where one of the decay products is missed.

The three mass spectra in Fig.~\ref{fig:fig2} present similar features.
\begin{itemize}
\item{}
A single bin peak at 2.4 \gevcc due to a reflection from the 
decays of the $D_{s1}(2536)^+$  
to $D^{*0} K^+$ or $D^{*+} K^0_S$ in which the $\pi^0$ or $\gamma$ from the 
$D^*$ decay is missed. This state, if $J^P=1^+$, cannot decay to $D K$.
\item{}
A prominent narrow signal due to the $D_{s2}(2573)^+$.
\item{}
A broad structure peaking at a mass of approximately 2.7~\gevcc.
\item{}
An enhancement around 2.86~\gevcc. This can be seen better in the expanded views
shown in the insets of Fig.~\ref{fig:fig2}.
\end{itemize}

In the following we examine different background sources: combinatorial, possible reflections
from $D^*$ decays, and particle misidentification. 

Backgrounds come both from events in which the candidate $D$ meson is 
correctly identified  and from events in which it is not.  
The first case can be studied combining
a reconstructed  $D$ meson with a kaon from another $\bar D$ meson in the same 
event, using data with 
fully reconstructed $D \bar D$ pairs or Monte Carlo simulations. No signal near 
2.7 or 2.86~\gevcc is seen in the $DK$ mass plots for these events.
The second case can be
studied using the $D$ mass sidebands.  The shaded regions in
Fig.~\ref{fig:fig2} show the $DK$ mass spectra for events in the
$D$ sideband regions normalized to the estimated background in the signal region. 
No prominent structure is visible in the sideband mass spectra.

We examined the possibility that the features at 2.7 and 2.86~\gevcc could be a
reflection from $D^*$ or other higher mass resonances.  
Candidate $DK$ pairs where the $D$ is a $D^*$-decay product are identified by 
forming $D\pi$ and $D\gamma$ combinations and requiring the invariant-mass difference 
between one of those combinations and the $D$ to be within $\pm 2\sigma$ of the known $D^*-D$ 
mass difference.
No signal near 2.7 or
2.86~\gevcc is seen in the $DK$ mass plots for these events.  
Events belonging to these
possible reflections (except for the $D^{*0} \to D^0 \gamma$ events, which could not be 
isolated cleanly) have been removed from 
the mass distributions shown
in Fig.~\ref{fig:fig2} (corresponding to $\approx$8\% of the final sample).
\begin{figure*}
\begin{center}
\vskip -0.15in
\includegraphics[width=1.0\linewidth]{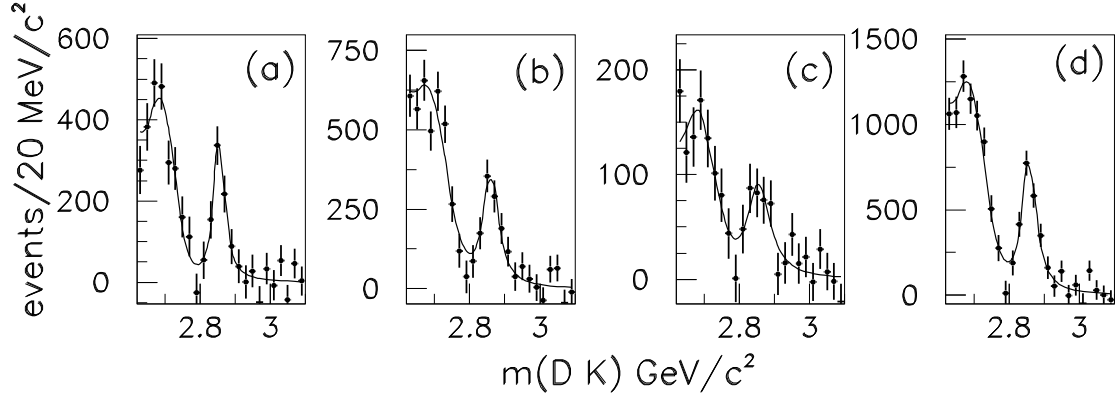}
\end{center}
\vskip -0.15in
\caption{\label{fig:fig3} Fitted background-subtracted $D K$ invariant mass 
distributions for 
(a) $D^0_{K^-\pi^+} K^+$, (b) $D^0_{K^-\pi^+\pi^0} K^+$, 
(c) $D^+_{K^-\pi^+\pi^+}K^0_s$, and (d) the sum of all modes in the 2.86 \gevcc mass region. 
The curves are the fitted functions described in the text.
}
\end{figure*}

\begin{table*}[tbp]
\caption{Results from the fits to the total $D K$ mass spectra of 
Fig.~\ref{fig:fig2}. Quantities are in
  units of \mevcc. Errors are statistical only. Simultaneous fits of the three
mass spectra are labelled with $DK_A$ and $DK_B$.}
\label{tab:fits}
\begin{center}
\vskip -0.2cm
\begin{tabular}{lccccccc}
\hline
Fit & $m(D_{s2}(2573)^+)$ & $\Gamma(D_{s2}(2573)^+)$ & $m(X(2690)^+)$ & $\sigma(X(2690)^+)$ & $m(D_{sJ}(2860)^+)$ & $\Gamma(D_{sJ}(2860)^+)$  &  $\chi^2/NDF$ \cr
\hline
$D^0_{K^-\pi^+} K^+$ & 2572.4 $\pm$ 0.4 &  27.6 $\pm$ 0.5 & 2687 $\pm$ 4 & 41 $\pm$ 5 & 2855.4 $\pm$ 2.0 & 37 $\pm$ 8 & 26/20 \cr
$D^0_{K^-\pi^+\pi^0} K^+$ & 2572.3 $\pm$ 0.5 &  28.4 $\pm$ 0.9 & 2682 $\pm$ 5 & 52 $\pm$ 5 & 2860.8 $\pm$ 4.0 & 52 $\pm$ 14 & 33/20 \cr
$D^+_{K^-\pi^+\pi^+} K^0_s$ & 2572.6 $\pm$ 0.9 &  21.9 $\pm$ 1.1 & 2684 $\pm$ 7 & 50 $\pm$ 7 & 2856.6 $\pm$ 8.0 & 81 $\pm$ 25 & 26/21 \cr
$DK_A$ & 2572.3 $\pm$ 0.3 &  27.1 $\pm$ 0.6 & 2684 $\pm$ 3 & 48 $\pm$ 2 & 2856.6 $\pm$ 1.5 & 47 $\pm$ 7 & 100/72 \cr
\hline
Fit & $m(D_{s2}(2573)^+)$ & $\Gamma(D_{s2}(2573)^+)$ & $m(X(2690)^+)$ & $\Gamma(X(2690)^+)$ & $m(D_{sJ}(2860)^+)$ & $\Gamma(D_{sJ}(2860)^+)$  &  $\chi^2/NDF$ \cr
\hline
$DK_B$& 2572.3 $\pm$ 0.3 &  27.0 $\pm$ 0.5 & 2688 $\pm$ 4 & 112 $\pm$ 7 & 2857.6 $\pm$ 1.9 & 38 $\pm$ 7 & 112/72 \cr
\hline
\end{tabular}
\end{center}
\end{table*}
 
We use a Monte Carlo simulation to investigate the possibility that 
the 2.7 or 2.86~\gevcc signals could 
be due to reflections from other charmed states.
This simulation includes $e^+e^-\to c\bar c$ events 
and all known charmed states and decays. The Monte Carlo events were
generated using a detailed 
detector simulation and subjected to the same reconstruction
and event-selection procedure as was used for the data. The $D^0 K^+$ 
effective mass distribution for these Monte Carlo events is shown (dotted)
in Fig.~\ref{fig:fig2}(a) for channel (1). The normalization is arbitrary.
No 
peak is found in the 2.7 and 2.86~\gevcc $D^0 K^+$ signal regions.
We note that the simulation underestimates the size of the $D_{s1}(2536)^+$ reflection and
the $D_{s2}(2573)^+$ signal relative to the background. No such discrepancy 
is found in the study of the $D^0 \pi^+$ final state, therefore we attribute 
this effect
to a poor knowledge of the strange-charmed meson cross sections. 
In order to improve the data-Monte Carlo comparison, the events having a
$D_{s1}(2536)^+$ and $D_{s2}(2573)^+$ in the final state have been scaled up 
by factors 5 and 2 respectively. 

We checked the possibility that the structures at 2.7 and 2.86~\gevcc are due to
misidentifying pions as kaons by assigning the kaon mass to the pion
in $D^0 \pi^+$  data events.  We observe no structure
near 2.7 or 2.86~\gevcc in the resulting $D^0K^+$ invariant mass distribution.
Monte Carlo simulations and tests using the data show that these structures 
also do not 
originate from protons misidentified as kaons from high mass charmed 
baryon decays.

Wrong sign $D^0 K^-$ mass distributions for channels (1) and (2) have also
been examined and we find no signal in either mass spectrum.

A more detailed study in channel (1)
of the 2.7 \gevcc structure shows a broad structure in this mass region
for events from the $D^0$ sidebands in which the $DK$ candidate has a very 
low $p^*$ ($p^* <$ 3 \gevc).
This is not seen in channels (2) or (3) however. We conclude that the 
assignment of 
the 2.7 \gevcc structure to a reflection remains inconclusive.

By comparing the reconstructed mass distributions for the $D K$ system
 with those generated with Monte Carlo simulations, we obtain the mass resolutions.
The resolutions are similar 
in the three channels, increasing linearly from 1.7 \mevcc
at a mass of 2.5 \gevcc, to 3.5  \mevcc at a mass of 2.86~\gevcc.

In the following discussion we label as $D_{sJ}(2860)^+$ the structure in the 2.86~\gevcc 
mass region and as $X(2690)^+$ the structure observed in the 2.7~\gevcc 
mass region. 
We fit to the three $D K$ mass spectra shown in Fig.~\ref{fig:fig2}
from 2.42 \gevcc to 3.1 \gevcc (excluding the $D_{s1}(2536)^+$ reflection)
using a binned $\chi^2$ minimization. The background for the three $DK$ mass
distributions is described by a threshold function: 
$(m - m_{\rm th})^{\alpha}~e^{-\beta m-\gamma m^2-\delta m^3}$ 
where $m_{\rm th}=m_{D}+m_K$. 
A fit to the Monte Carlo distribution shown in 
Fig.~\ref{fig:fig2}(a) using this background expression and one spin-2 
relativistic Breit-Wigner 
for the $D_{s2}(2573)^+$ gives a good fit with 32 \% $\chi^2$ probability. 
In the fit to the data, the $D_{s2}(2573)^+$ and $D_{sJ}(2860)^+$ peaks are
described with
relativistic Breit-Wigner lineshapes where spin-2 is assumed for the $D_{s2}(2573)^+$ and spin 0 
is used for the $D_{sJ}(2860)^+$. 
We find that the $D_{sJ}(2860)^+$ parameters
are insensitive to the choice of the spin. 
The best description
of the $X(2690)^+$ structure is obtained using a Gaussian distribution.
The results from the fits are summarized in Table~\ref{tab:fits}.
Table~\ref{tab:sing} summarizes the $\chi^2$ probabilities, the number of $D_{sJ}(2860)^+$ 
events (with statistical and systematic errors) and the $D_{sJ}(2860)^+$ statistical 
significances 
from the three separate fits to the $DK$ mass spectra.

\begin{table}[tbp]
\caption{$\chi^2$ probabilities, $D_{sJ}(2860)^+$ event yields and statistical significances from 
the three separate fits to the total $DK$ mass spectra of 
Fig.~\ref{fig:fig2}.}
\label{tab:sing}
\begin{center}
\vskip -0.2cm
\begin{tabular}{lccc}
\hline
Channel & $\chi^2$ & $D_{sJ}(2860)^+$  &  Statistical \cr
        & probability (\%)& events &  significance  \cr
\hline
$D^0_{K^-\pi^+} K^+$ & 17 & 886 $\pm$ 134 $\pm$ 49 & 6.2 $\sigma$ \cr
$D^0_{K^-\pi^+\pi^0} K^+$ & 3  & 1146 $\pm$ 157 $\pm$ 78 & 6.5 $\sigma$ \cr
$D^+_{K^-\pi^+\pi^+} K^0_s$ & 21 & 371 $\pm$ 84 $\pm$ 53 & 3.7 $\sigma$ \cr
$DK_A$ & 1.6 & 2717 $\pm$ 262 $\pm$ 190 & 8.4 $\sigma$ \cr
$DK_B$ & 0.2 & 2161 $\pm$ 238 $\pm$ 151 & 7.7 $\sigma$ \cr
\hline
\end{tabular}
\end{center}
\end{table}

The fits give  
consistent values for the parameters of the three structures. We notice a smaller width
of the $D_{s2}(2573)^+$ in the $D^+_{K^-\pi^+\pi^+} K^0_s$ channel 
which we attribute to the 
uncertainty in the description of the background.
We compute also the ratios
of the yields of $D_{sJ}(2860)^+$ with respect to $D_{s2}(2573)^+$ finding agreement, 
within statistical errors, between the three channels.  

The presence of resonant structures can be visually enhanced by subtracting the
fitted background threshold function from the data.
Fig.~\ref{fig:fig3} shows the background-subtracted $D^0_{K^-\pi^+} K^+$,
$D^0_{K^-\pi^+\pi^0} K^+$, and $D^+_{K^-\pi^+\pi^+} K^0_s$
invariant mass 
distributions in the 2.86 \gevcc mass region. 
Fig.~\ref{fig:fig3}(d) shows the sum
of the three mass spectra.

We also fit to the three distributions simultaneously. The parameters from this fit are 
labelled $DK_A$ in Table~\ref{tab:fits}.
If we remove the $D_{sJ}(2860)^+$, the $\chi^2$ increases by
108 units while the number of degrees of freedom increases by five.  

As a
systematic test, we repeated the fits varying the lower $p^*$ cut on the $DK$ system  
from 3.50 to 3.75 and to 4.00 \gevc. 
We also restricted the fit to the $D_{sJ}(2860)^+$ only and replaced the threshold function which
represents the background with a polynomial. Fits have also been performed without removing the 
events associated to $D^*$ reflections and modifying the spin of $D_{s2}(2573)^+$. 
The
systematic uncertainties take into account the variation of the resonance 
parameters among the three different final states and the resonance parameterizations. 
The uncertainty on the mass
scale is estimated to be of the order of 1 \mevcc.

We obtain the mass and 
width of $D_{s2}(2573)^+$:
\begin{eqnarray*}
m(D_{s2}(2573)^+) = (2572.2 \pm 0.3 \pm 1.0) \ \mevcc \\
\Gamma(D_{s2}(2573)^+)=(27.1 \pm 0.6 \pm 5.6) \ \mevcc, 
\end{eqnarray*}
where the first errors are statistical and the second systematic. 
For the new state we find
\begin{eqnarray*}
m(D_{sJ}(2860)^+) = (2856.6 \pm 1.5 \pm 5.0) \ \mevcc \\
\Gamma(D_{sJ}(2860)^+)=(47 \pm 7 \pm 10) \ \mevcc.
\end{eqnarray*}
Since the assignment of the $X(2690)^+$ as a reflection is inconclusive, 
the  three mass  spectra have  also been  fit including 
the $X(2690)^+$ as 
an additional resonance (Breit-Wigner, rather than Gaussian shape). 
This gives fit $DK_B$ shown  in Table~\ref{tab:fits}. 
The resulting resonance parameters are:
\begin{eqnarray*}
m(X(2690)^+)= (2688 \pm 4  \pm 3) \  \mevcc \\
\Gamma(X(2690)^+)=(112 \pm 7 \pm 36) \ \mevcc.
\end{eqnarray*}

In summary, in 240~${\rm fb}^{-1}$ of data collected
by the \babar\  experiment, 
we observe a new $D_s^+$  state in the inclusive $D K$
mass distribution near 2.86~\gevcc in three independent channels. 
The decay to two pseudoscalar mesons implies
a natural spin-parity for this state: $J^P=0^+, 1^-,\dots$. 
It has been suggested that this new state 
could be a radial excitation of $D^*_{sJ}(2317)$~\cite{beveren} although
other possibilities cannot be ruled out.
In the same mass distributions we also observe a broad enhancement around 2.69 \gevcc
which it is not possible to associate to any known reflection or background.

We are grateful for the excellent luminosity and machine conditions
provided by our \pep2\ colleagues, 
and for the substantial dedicated effort from
the computing organizations that support \babar.
The collaborating institutions wish to thank 
SLAC for its support and kind hospitality. 
This work is supported by
DOE
and NSF (USA),
NSERC (Canada),
IHEP (China),
CEA and
CNRS-IN2P3
(France),
BMBF and DFG
(Germany),
INFN (Italy),
FOM (The Netherlands),
NFR (Norway),
MIST (Russia),
MEC (Spain), and
PPARC (United Kingdom). 
Individuals have received support from the
Marie Curie EIF (European Union) and
the A.~P.~Sloan Foundation.

\bibliography{note616}

\begin{thebibliography}{99}
\bibitem{footnote}
Charge-conjugate reactions are implied throughout.
\end{thebibliography}

\end{document}